# Earth's Inner Core Periodic Motion due to Pressure Difference Induced by Tidal Acceleration

Martin Wolf                                  E-mail:emartin.wolf@gmail.com

*Summary*

The inner structure of the earth is still a topic of discussion. Seismic measurements showed a structure of solid, liquid, solid which describes the mantle, outer core and inner core with the inner core in the center. The analysis of waveform doublets suggests now that the inner core is out of center and even of faster rotation than the mantel and crust. From the sum of Buoyancy and Gravity on the earth's inner core, its position energy is plotted and together with the tangential tidal acceleration, it is derived that Earth´s Inner Core cannot be in a center position without an additional force. The Earth Core System is explained as Hydrodynamic Bearing. The tidal acceleration is identified as the reason for the periodic motion of the inner core and certain frequencies of nutation. The Eccentricities responsible for nutation due to the effects from the sun and moon are calculated as an approximation.

1. **Buoyancy and Gravity**

To investigate the inner earth, the forces on the Earth Inner Core (EIC) are calculated here, using PREM data.

It was found to be useful to calculate the Buoyancy (**B**) and the Gravity (**G**) on the EIC as a function of eccentricity (E) in a numerical way. The centrifugal force due to the initial rotation of the earth is not considered here because the influence is minor.

With the EIC out of center, there is a **G** on the EIC coming from the liquid outer core surrounding EIC. The liquid mass adding up to **G** has to be within the smallest globe concentric to the earth's center, including EIC (Figure 1), whose radius is given by r.

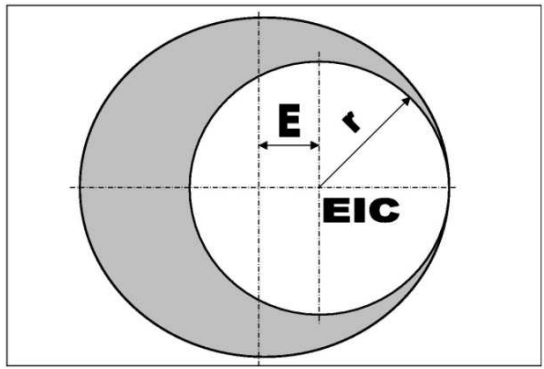

**Figure 1: The grey area marks the mass responsible for the gravity on the Earth's inner core (schematically).**

If the EIC is in the center, the forces from pressure from all the sides cancel out each other, therefore the **B** is zero (Figure 2). With growing E, **B** rises (Figure 2).

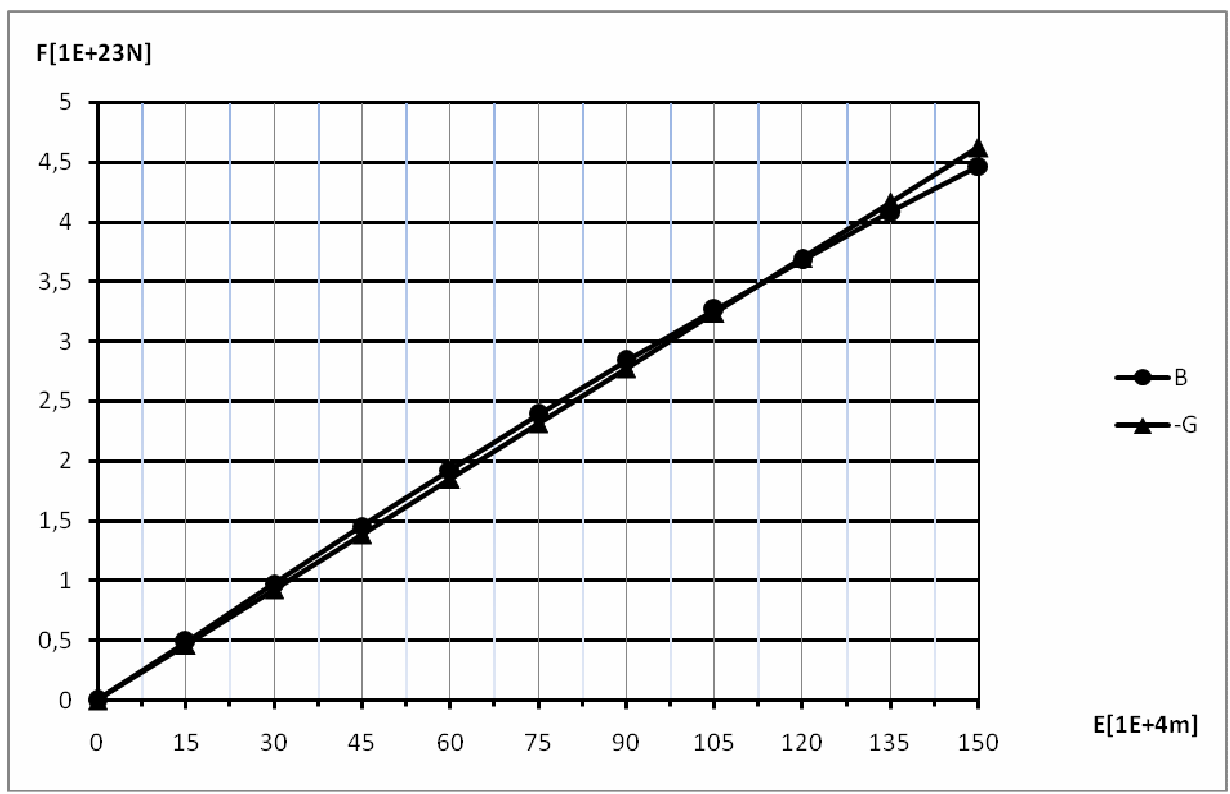

**Figure 2**: **The buoyancy and the minus of gravity are equal at eccentricity zero and near 1150 km.**

From Figure 2 it can be seen that there are several positions where the sum of forces on the EIC is zero. **B** dominates over **G** near the center.

To be able to judge the stability of EIC the position energy (U) of the EIC out of the sum of **B** and **G**, has to be examined.

In Figure 3 it becomes clear that there is an unstable equilibrium for the EIC at eccentricity zero.

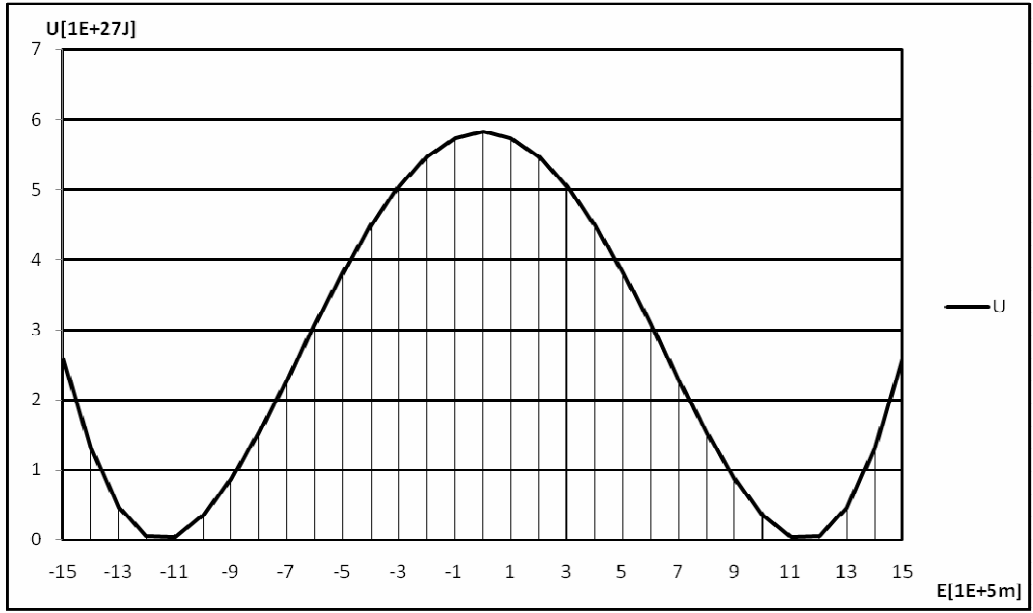

**Figure 3:** The **Position energy U of the earth inner core out of the sum of buoyancy and gravity over the eccentricity**.

Out of the center EIC would be driven away from the center. The equilibrium at about +/- 1150 km would be stable.

## 2. Tidal effect

Now the Tangential Tidal effect has to be taken into account. The sum of centrifugal acceleration ($a_c$) and gravitational acceleration ($a_g$) produce the tidal effect. Due to the eccentricity of the earth orbiting the sun, there is a strong tangential tidal acceleration ($a_t$) on the earth at the position C (Figure 4). The magnitude can be calculated as an approximation by:

$$a_t(C) = \varepsilon\, m\, \gamma\, /\, a^2. \tag{2.1}$$

$\varepsilon$ is the numeric eccentricity (in this case of the earth orbit) equal to $\sin(\alpha)$, with $\alpha$ as in Figure 4, m is the mass of the sun in this case, $\gamma$ is the gravitational constant and **a** is the distance F1 to C in this case sun earth (Figure 4).

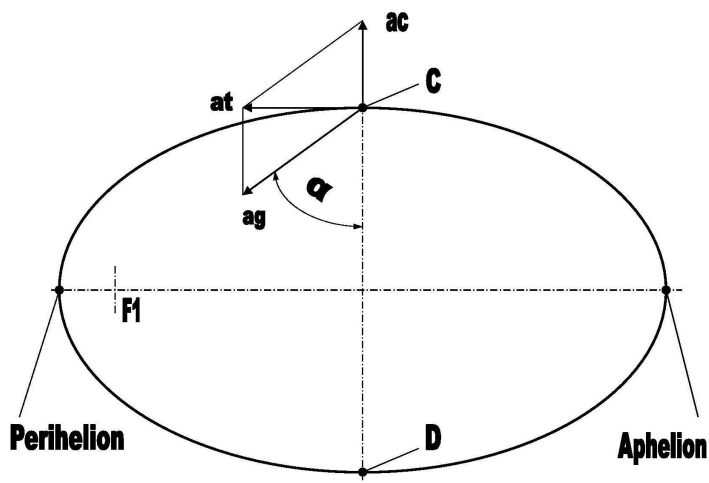

**Figure 4: The Earth's tangential tidal acceleration (schematically).**

The peak value of $a_t$ appears very close to C on the Perihelion side and despite the fact that $a_t(C)$ according to 2.1 is an approximation, it is almost identically to the value at the peak of the sun earth system:

$a_t(C)$ = 9.91E-05 m/s² .

In an anomalistic period there is a non sinusoidal periodic behavior with two peaks per year near C and D (Figure 4). Due to the axial tilt of the earth $a_t$ can be separated into an equatorial (**eq**) and a polar (**po**) part (Figure 5). Superposing $a_t$ over to the gravitational acceleration (**g**) produces:

**g'** = **$a_t$** + **g**.                                                                                                                              (2.2)

Since **g** is a vector pointing to the center of the earth and $a_t$ can be considered as a constant vector over the earth core in this context, **g'** is increased by $a_t$ on one hemisphere and diminished on the other he hemisphere. The position of **g'** equal to zero is 27.5 m eccentric to the Perihelion side at maxima of $a_t$.

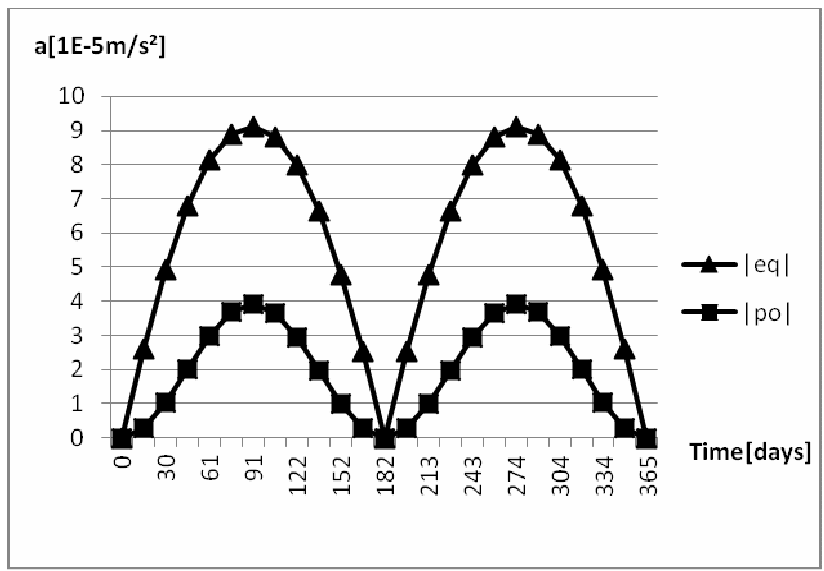

**Figure 5: The absolute value of the tangential tidal acceleration from the sun to the earth in an anomalistic year separated in its equatorial (eq) and polar (po) part.**

It is more interesting that the hydrostatic pressure (p) at the EIC also changes:

p = Σ ρ **g'** h.                                                                                                                                       (2.3)

ρ is the density of the layer, **g'** is the average total acceleration at the corresponding layer and h is the height of the layer. The liquid hemisphere on one side is periodic subject to a higher **g'** than the one on other side and subsequently the hydrostatic pressure also changes.

With an EIC in an unstable equilibrium at center position plus a periodic changing hydrostatic pressure situation on the EIC, it should be underlined that the EIC in center or near center position [5] cannot be explained by **B**, **G** and $a_t$. The model described so far is theoretical because EIC would drift away to the position of E near 1150 km. There is a force missing that keeps EIC concentric or near concentric.

3. **Hydrodynamic Effect**

Now it has to be remembered that the earth is rotating with an approximately constant angular velocity. The external asymmetric influence of $a_t$ together with **B** ensures that the EIC remains in the Perihelion hemisphere. A rotating earth with an EIC which stays eccentric towards the Perihelion side forces the liquid outer core to flow around an obstacle. The liquid outer core passes a nozzle like geometry towards the point of minimum distance between EIC and mantel (Figure 6). This produces a hydrodynamic force with the necessary stabilizing quality that keeps the drifting of the EIC in limits. The given system is what in technical terms is called a Hydrodynamic Bearing (HB). The force out of such a HB depends on (a) the relative velocity ($v_r$) between the moving partners (EIC and solid mantel), (b) the dynamic viscosity of the liquid between the partners subject to shear velocity, (c) the charged area and (d) the eccentricity that creates the nozzle like geometry [3]. The $v_r$ between EIC and the boundary layer is a consequence of the supposed constant angular velocity (Figure 6).

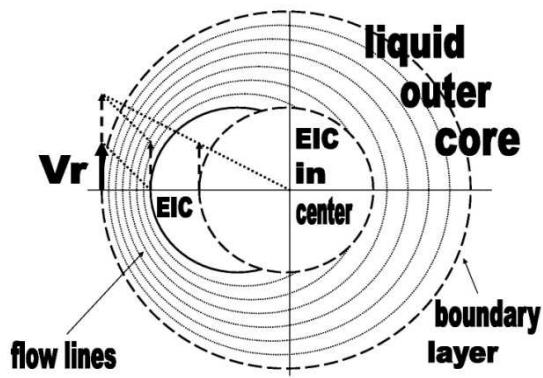

**Figure 6: The relative velocity between EIC and the boundary layer in the plane of motion, plus the flow lines for the supposed laminar flow, given schematically.**

The stream of liquid material passing the EIC gives an accelerating torsional momentum to it (the partner with the lower velocity). The EIC has been slightly accelerated by this periodic momentum over a long period of time. Therefore, seismic measurements proposing a faster rotating EIC [6] are proved to be accurate by this work.

4. **Earth as a Gyroscope**

The motion of the inner core raises the question: How much is the displacement? From astronomic observations it is known that the earth is subject to a motion called nutation of the same frequency as $a_t$. To do the exact calculation to work out the eccentricity of the EIC from certain frequencies of nutation would involve solving the Liouville's theorem [2] with torsional momentum zero [1]:

$$\frac{d}{dt}[J(t)\omega(t) + l(t)] + \omega(t) \times [J(t)\omega(t) + l(t)] = 0 \qquad (4.1)$$

ω being the angular velocity, **l** the vector of relative angular momentum and J the mass momentum of inertia.

Assuming that only the **ω** and J of the rotation axis of earth are changing permits an approximation. The movement of the EIC could be calculated because of:

$$\frac{d}{dt} L = 0 \qquad (4.2)$$

and

**L** = J ω.  (4.3)

**L** is the angular momentum vector. The change in **ω** equal to an earth rotation variation can be described by the nutation of length of 1.3187" with a period of 182.6 days [4] consistent with $a_t$ in frequency. The eccentricity necessary is about 40 km at maximum due to the influence of the sun. There is also a nutation of length of 0.2274" with a period of 13.7 days [4] consistent with the influence of the moon. The result is an eccentricity of about 60 km to produce this nutation. These values are small compared to the diameter of EIC of 2443 km. For a HB the eccentricities are especially small and a HB on such low eccentricities is known to be susceptible to oscillations [3].

5. Conclusion

The detail that **B** dominates over **G** near the center is the main cause for the fact that EIC cannot be in the center. The $a_t$ is the trigger for the movement out of center, also providing its direction and giving a periodic character to the movement. The main movement of EIC is in direction of **eq** provoking the hydrodynamic counter force. The movement in the direction of **po** is expected to be smaller because **po** is smaller than **eq**. The hydrodynamic effect in polar direction is also smaller since there is no rotational velocity at the pole and therefore no relative velocity. Therefore, the movement in polar direction might be disproportionately larger than expected. The axial tilt of earth and the direction of $a_t$ are in the same orientation during a full rotation such that **po** aims at the northern pole. Therefore, the polar eccentric movement of EIC is mainly to the northern hemisphere. The movement in polar direction was already noticed by the analysis of waveform doublets [5]. The author suggests conducting the analysis of waveform doublets in the axis Colombia – Singapore or similar to measure the equatorial movement of EIC.